\documentclass[12pt,titlepage]{article}
\usepackage{setspace}
\usepackage{amsmath}
\usepackage[letterpaper,top=1in,bottom=1in,left=1in,right=1in]{geometry}
 \usepackage{gensymb}
\usepackage{graphicx}
\usepackage{times}

\usepackage[round,numbers,sort&compress]{natbib}

\newcommand{\be}{\begin{equation}}
\newcommand{\ee}{\end{equation}}
\newcommand{\bea}{\begin{eqnarray*}}
\newcommand{\eea}{\end{eqnarray*}}
\newcommand{\beal}{\begin{eqnarray}}
\newcommand{\eeal}{\end{eqnarray}}

\title{Arrangement of Annexin II tetramer and its impact on the structure and diffusivity of supported lipid bilayers}
\author{Kirstin Fritz \\
    Department f\"{u}r Physik und CeNS,
    Ludwig-Maximilians-Universit\"{a}t, \\
    M\"{u}nchen, Germany
    \and Georg Fritz\\
    Arnold Sommerfeld Center for Theoretical Physics und CeNS,\\
    Ludwig-Maximilians-Universit\"{a}t, \\
     M\"{u}nchen, Germany
    \and Barbara Windschiegl\\
    Institut f\"ur Organische und Biomolekulare Chemie, Georg August Universit\"at \\
    G\"ottingen, Germany \\
    \and Claudia Steinem\\
    Institut f\"ur Organische und Biomolekulare Chemie, Georg August Universit\"at \\
    G\"ottingen, Germany \\
    \and Bert Nickel\thanks{
           Corresponding author. Full address:
           Department f\"{u}r Physik and CeNS,
       Ludwig-Maximilians-Universit\"{a}t,
       Geschwister-Scholl-Platz 1,
       D-80539 M\"{u}nchen, Germany,
       Tel.:~(+49)~89~2180-1460, Fax:~(+49)~89~2180-1452, E-mail: nickel@lmu.de}\\
    Department f\"{u}r Physik und CeNS,
    Ludwig-Maximilians-Universit\"{a}t,\\
    M\"{u}nchen, Germany}

\begin{document}

\maketitle
\abstract{Annexins are a family of proteins that bind to anionic phospholipid membranes in a $\mathrm{Ca^{2+}}$-dependent manner. Annexin A2 forms heterotetramers (Anx A2t) with the S100A10 (p11) protein dimer. The tetramer is capable of bridging phospholipid membranes and it has been
suggested to play a role in $\mathrm{Ca}^{2+}$-dependent exocytosis and cell-cell adhesion of metastatic cells. Here, we employ x-ray reflectivity measurements to resolve the conformation of Anx A2t upon $\mathrm{Ca}^{2+}$-dependent binding to single supported lipid bilayers (SLBs) composed of different mixtures of anionic (POPS) and neutral (POPC) phospholipids. Based on our results  we propose that Anx A2t binds in a side-by-side configuration, {\it i.e.}, both Anx A2 monomers bind to the bilayer with the p11 dimer positioned on top. Furthermore, we observe a strong decrease of lipid mobility upon binding of Anx A2t to SLBs with varying POPS content. X-ray reflectivity measurements indicate that binding of Anx A2t also increases the density of the SLB. Interestingly, in the protein-facing leaflet of the SLB the lipid density is higher than in the substrate-facing leaflet. This asymmetric densification of the lipid bilayer by Anx A2t and $\mathrm{Ca}^{2+}$ might have important implications for the biochemical mechanism of Anx A2t-induced endo- and exocytosis.
}
\clearpage

\section*{Introduction}

Annexin A2 (Anx A2) is a $\mathrm{Ca}^{2+}$-binding protein which binds to acidic phospholipids and is involved in many cellular regulatory processes, such as the regulation of vesicular trafficking, endosome fusion, insulin signal transduction \cite{Biener_JBiolChem_96} and RNA binding \cite{Filipenko_JBiolChem_04}. Like other members of the Annexin protein family, it consists of two domains: the conserved core domain harboring the  $\mathrm{Ca}^{2+}$-binding sites and, among different Annexins, a variable N-terminal domain exposing interaction sites for other protein partners. In particular, Anx A2 forms a heterotetrameric complex (Anx A2t) with S100A10 (p11). S100A10 belongs to the S100 protein family, although it is distinct from the other members of this family as it does not undergo $\mathrm{Ca^{2+}}$-dependent conformational changes. Even in the absence of  $\mathrm{Ca}^{2+}$, S100A10 is in the active state and, like most other S100 proteins, forms an anti-parallelly packed non-covalent homodimer \cite{Rety_NatStructBiol_99}.

\begin{figure}
\centerline{\includegraphics[width=6in]{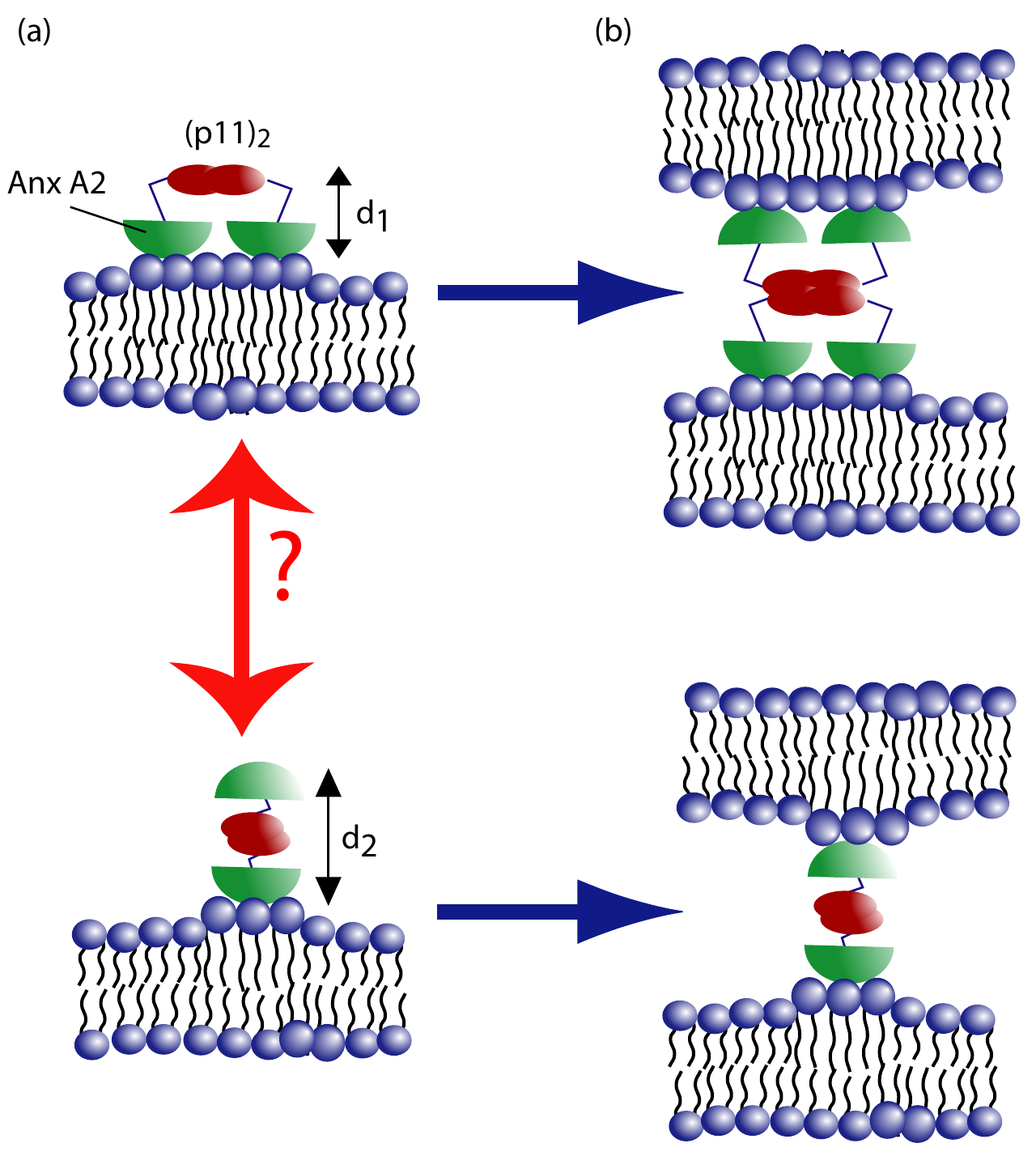}}
\caption{\label{FIGscheme} Fritz  et al.}
\end{figure}

To date, no experimentally resolved high-resolution structure of the full length complex of Anx A2t is available. However, data of the Anx A2 monomer missing the first 19 amino acids \cite{Burger_JMolBiol_96,Rosengarth_Annexins_04}, as well as of the complex between dimeric p11 and two synthetic N-terminal fragments composed of the first 11 amino acids of Anx A2 exist  \cite{Rety_NatStructBiol_99}. Based on these data, a structural model of the Anx A2t complex has been created by computational modeling \cite{Sopkova-deOliveiraSantos_BiochimBiophysActa_00}. Yet, there is still a controversial discussion about the organization of Anx A2t when bound to a single membrane and the complex that is formed upon membrane-membrane connection. On the one hand, when binding to a single surface supported membrane, the thickness of an Anx A2t layer obtained from scanning force microscopy experiments suggested that two Anx A2 monomers are connected by a p11 dimer and are bound in a ``side-by-side'' configuration to the membrane interface \cite{Menke_ChemBioChem_04}, see Fig.~\ref{FIGscheme} ({\it a, top}).  It has frequently been proposed that the contact between membranes may then be mediated via the interaction of two opposing Anx A2t complexes by formation of a heterooctameric structure, see Fig.~\ref{FIGscheme}~({\it b, top}) \cite{Lewit-Bentley_CellBiolInt_00, Menke_ChemBioChem_04,Schulz_Proteins_07}. On the other hand, based on cryo-electron microscopy results on Anx A2t-connecting vesicles \cite{Lambert_JMolBiol_97}, it was proposed that the Anx A2t complex bridges membranes in a ``vertical'' configuration, in which each Anx A2 monomer binds to one of the membranes while the p11 dimer is located in between \cite{Sopkova-deOliveiraSantos_BiochimBiophysActa_00}, see Fig.~\ref{FIGscheme}~({\it b, bottom}). From these measurements the same, albeit hypothetical, structure was proposed for the protein when binding to a single bilayer \cite{Menke_ChemBioChem_04}, see Fig.~\ref{FIGscheme}~({\it a, bottom}). Importantly, the vertical conformation of Annexin A2t, as obtained from membrane-membrane junctions, is predicted to result in a larger thickness of the protein layer than the side-by-side conformation \cite{Nakata_JCellBiol_90,Sopkova-deOliveiraSantos_BiochimBiophysActa_00,Lambert_JBiolChem_04,Schulz_Proteins_07}. However, during scanning force microscopy experiments as in ref.~\cite{Menke_ChemBioChem_04} a certain force is exerted on the soft protein layer and therefore its thickness might be underestimated. Hence, to date, it cannot be excluded that Anx A2t also binds to a single lipid bilayer in the vertical conformation.

In this paper we used x-ray reflectivity (XR) to investigate the conformation of Anx A2t and its impact on the structure of single supported lipid bilayers (SLBs). In the field of SLBs, XR has been successfully applied as a non-invasive tool to resolve subnano-structural features \cite{Miller_PhysRevLett_05,Reich_RevSciInstrum_05,Nickel_Biointerphases_08}. Differences between gel and fluid lipid bilayers have been characterized \cite{Novakova_PhysRevE_06,Reich_BiophysJ_08} in progress toward understanding heterogeneous lipid mixtures. In contrast to neutron scattering techniques, XR measurements using modern synchrotron sources achieve a superior resolution. However, due to the low contrast only few XR measurements have been conducted on protein binding to SLBs \cite{Horton_Langmuir_07}. Here we demonstrate that a single Anx A2t layer provides enough contrast for a high-resolution XR study of the protein-SLB structure. It enables us to elucidate the thickness of Anx A2t bound to a single bilayer and to discriminate between the side-by-side and vertical configuration. A major advantage of XR is that also the structure of the covered SLB, {\it i.e.}, the head-to-head distance and the packing density of the lipids, remains experimentally accessible upon protein binding. Hence, the influence of protein binding and substrate on the bilayer structure can be monitored. In particular, it has been proposed that binding of Anx A2t creates pores in the membrane through densification of the bilayer \cite{Chasserot2005} and thereby exerts its role in endo- and exocytosis. However, only few measurements investigate the influence of Anx A2t binding on the structure and density of the bilayer. In this paper a careful determination of the electron density profiles is reported and further decomposition into components of the bilayer allow us to analyze properties of both bilayer leaflets independently, so that differences in packing density can be resolved. By using different ratios of anionic POPS to neutral POPC we examine the influence of Anx A2t binding on these structural bilayer parameters.  Furthermore, it was hypothesized that the protein collects negatively charged lipids below itself and thus may alter the fluidity and state of the bilayer \cite{ross2005}. To test this conjecture we complement our x-ray studies by continuous bleaching measurements on fluorescently labeled lipid probes. The combination of both experimental methods allows us to correlate the structural rearrangement induced by protein binding to changes in the diffusion properties of the lipid bilayer. Last,   possible implications for endo- and exocytotic processes are discussed.

\section*{Materials and Methods}

\footnotesize

\paragraph{\footnotesize SLB formation.} Surface supported bilayers of varying molar composition of 1-palmitoyl-2-oleoyl-{\it sn}-glycero-3-phosphocholine (POPC), and 1-palmitoyl-2-oleoyl-{\it sn}-glycero-3-[phospho-L-serine] (sodium salt) (POPS), were prepared on silicon oxide by the following procedure: Appropriate amounts of lipids were dissolved in chloroform and filled in a clean glass vial. 0.5 mol \% fluorescent dye (Texas Red 1,2-dihexadecanoyl-{\it sn}-glycero-3-phosphoethanolamine, triethylammonium salt, Texas Red DHPE or Oregon Green 488 1,2-dihexadecanoyl-{\it sn}-glycero-3-phosphoethanolamine, Oregon Green 488 DHPE, Invitrogen, Germany) were added to the mixture for fluorescence microscopy. The solvent was evaporated by a nitrogen flow. After removal of residual solvent overnight in a dessicator connected to a rotary vacuum pump, the mixture was dispersed in a PBS buffer (Sigma-Aldrich) with 1 M NaCl to a lipid concentration of 1 mg/ml and vortexed to form a lipid suspension. The suspension was kept at $40 \celsius$ for two hours. Large unilamellar vesicles were obtained by extrusion through 100 nm polycarbonate filters using an extruder (Avestin, Ottawa, Canada). 150\,$\mu$l of the vesicle suspension were injected into the microfluidic chamber, which holds the silicon wafer. For a detailed description of the microfluidic chamber please refer to ref. \cite{Reich_RevSciInstrum_05}. Silicon wafers with thermal oxide layers were bought from Crystec (Berlin, Germany) and cut to a sample size of 15x20 $\mathrm{mm^2}$.  To avoid fluorescence quenching the thickness of the oxide layer on the silicon surfaces ranged from 50 nm to 100 nm. Prior to usage, silicon wafers were cleaned chemically by a three stage protocol \cite{Reich_RevSciInstrum_05,Reich_BiophysJ_08} involving $\mathrm{H_2O_2/HCl/H_2O}$ and $\mathrm{H_2O_2/NH_3/H_2O}$. They were stored in deionized water and used within 24\,h. The bare silicon surface was measured by x-ray reflectivity and the obtained roughness of approximately $4\, \AA$ was fixed during the decomposition of the electron density profile. The sample was incubated for at least three hours at room temperature, then rinsed intensively with buffer to remove excess vesicles, and monitored by fluorescence microscopy for homogeneity. In a following step, the buffer was carefully exchanged by deionized pure water (MilliQ, specific resistivity 18.2 M$\Omega$/cm, Millipore Corp. Billerica, Massachusetts), which was injected into one of the two outlets. This induced an osmotic shock promoting fusion of the adsorbed vesicles to the substrate and resulted in SLB formation. The SLB coverage was controlled by fluorescence microscopy. When homogenous coverage was reached, 20 mM Tris/HCl, 100 mM NaCl, 1 mM $\mathrm{CaCl_2}$, 0.5 mM DTT, pH 7.4 was injected.

\paragraph{\footnotesize Protein binding and unbinding.} Annexin A2 tetramer was extracted from bovine intestines as described previously \cite{Gerke_EMBOJ_84}. The protein was stored at $-20 \celsius$ in 20 mM MES, 100 mM NaCl, 0.5 mM EGTA, pH 6.0. Before usage, the protein solution was carefully warmed up to $4\celsius$ in an ice bath and 1 ml of 20 mM Tris/HCl, 100 mM NaCl, 1 mM $\mathrm{CaCl_2}$, 0.5 mM DTT, pH 7.4 was added to the protein solution. The solution was further diluted to a concentration of 3-6\,$\mu$M. $150\, \mu l$ of protein solution were then added to the SLB in the microfluidic chamber. Bilayers were incubated for at least three hours at room temperature. After incubation, the microfluidic chamber was rinsed with 20 mM Tris/HCl, 100 mM NaCl, 1 mM $\mathrm{CaCl_2}$, 0.5 mM DTT, pH 7.4 to remove excess protein. To unbind Anx A2t completely from the membrane, the sample was rinsed with 20 mM Tris/HCl, 100 mM NaCl, 10 mM EDTA, pH 7.4.

\paragraph{\footnotesize Fluorescence microscopy.} SLBs were controlled by fluorescence microscopy on site at HASYLAB and ESRF using a transportable Zeiss Axiotech vario fluorescence microscope (Oberkochen, Germany) equipped with 10$\times$ (NA 0.3) and long-distance 63$\times$ (NA 0.75) Plan-Neofluar objectives. Images were captured with an ORCA C4742-95 CCD camera and WASABI imaging software from Hamamatsu Photonics (Tutzing, Germany). For continuous bleaching experiments, a 120 W mercury short arc reflector lamp (HXP-R120W) was used. Continuous bleaching data were analyzed using self-written code based on MATLAB, The MathWorks, Inc..

\paragraph{\footnotesize X-ray reflectivity.} X-ray reflectivity measurements were performed at the beamline D4 at the Hamburger Synchrotronstrahlungslabor (HASYLAB) at the Deutsches Elektronen Synchrotron (DESY), Hamburg, Germany, and at the beamline ID01 at the European Synchrotron Radiation Facility (ESRF) in Grenoble, France. An x-ray energy of 19.75 keV was chosen to maximize the reflectivity signal while minimizing the beam damage in the microfluidic chambers. \cite{Reich_RevSciInstrum_05}.  Sample chambers were mounted in a horizontal scattering geometry as described previously \cite{Reich_RevSciInstrum_05,Horton_Langmuir_07}. Briefly, the incident beam enters the microfluidic chamber through a topas foil and passes through a $200\,\mu$m water-filled channel before hitting the sample.  Reflected intensities were collected by tilting the sample between incident angles  $\mathrm \theta$ = 0.02 and 2 degrees with a step number 142 at ESRF. For data collected at HASYLAB the sample was tilted between 0.026 and 1.51 degrees with a step number of 243 points. This leads to a momentum transfer $q$ normal to the surface up to $q = 0.53 \,\AA^{-1}$ at HASYLAB and $q= 0.71\,\AA^{-1}$ at ESRF. Here $q$ is defined by $q=2\pi\sin(\theta)/ \lambda$, where $\lambda$ is the wavelength of the incoming beam and $\theta$ the incident angle. The beam cross section was defined by a presample aperture of 80 $\mu$m horizontal and 100 $\mu$m vertical at D4 and 250 $\mu$m horizontal and 1000 $\mu$m vertical at ID01. Evacuated beam guides with Kapton windows positioned close to the sample chamber minimized air scattering. The reflected intensity was collected with a NaI (cyberstar) detector. For each data point the reflected intensity was collected for one second.  Automatic attenuators in front of the sample were used in order to reduce exposure to the full beam intensity. Furthermore, the sample was protected by a fast shutter system during motor movement. To control for radiation damage of bilayer and protein the samples were measured at the same sample position for a second time. No change in reflectivity signal of bilayer and protein layer was observed during this procedure. By detuning from the reflection condition, the background was determined and subtracted. The remaining reflection signal was corrected for illumination (footprint correction) and normalized to a reflectivity of one in the total reflection region. For graphical presentation, the data were multiplied by $q^4$ , to compensate for the overall decay of the reflectivity signal.

\paragraph{\footnotesize X-ray data analysis.} The reflectivity data were fitted with Parrats algorithm \cite{Parratt_PhysRev_54,Nevot_RevPhysAppl_80}. Two different fitting strategies were performed; in a first approach a small number of layers were chosen, each layer representing different regions of the bilayer \cite{Mattai_Biochemistry_89, Reich_BiophysJ_08}. Layer thickness and electron density were chosen as fitting parameters. In a second approach, a larger number of layers was created (10-12 layers) and only the electron density of each layer was used as fitting parameter. In this case, the thickness of each layer was set to the minimal layer thickness experimentally accessible (5-7 $\mathrm{\AA}$, see below) \cite{Daniel_Biointerphases_07}. The layer roughness was set to zero in both approaches. Within the first approach, the electron density of the bilayer was calculated with two slabs for each headgroup \cite{Mattai_Biochemistry_89, Reich_BiophysJ_08}, and three for the hydrophobic core of the bilayer \cite{Novakova_PhysRevE_06}, for measurements performed at ID01. Due to the lower resolution at D4, the bilayer data from this instrument were fitted with only one slab for each headgroup and three slabs for the hydrophobic core of the bilayer. In both measurements, an additional layer was added to account for the substrateÕs roughness and hydration of the headgroup proximal to the silicon substrate.
Within the second approach, the layer thickness was chosen to match the resolution of the experimental data. The resolution in a reflectometry experiment can be estimated by ($\pi/q_{max}$) \cite{Daniel_Biointerphases_07}. Here $q_{max}$ represents the maximum $q$ value achieved before the reflectivity signal decreases below the background signal. This leads to a slab thickness of 7 $\mathrm{\AA}$ and 5 $\mathrm{\AA}$ for data obtained at D4 and ID01, respectively. Thirteen and ten layers representing the bilayer were chosen for data from ID01 and D4, respectively. Further layers did not improve the quality of the fit. An average electron density was calculated from the fifteen best fits of both models. To further smoothen the profile a running average with a window size corresponding to the resolution of the experiments was applied. A similar approach was chosen to extract the electron density profile of the SLB covered by the protein layer.

\paragraph{\footnotesize Decomposition of electron density profiles.} Electron density profiles were decomposed into several component groups as described in the main text. The parameters of the decomposition were estimated by using a trust-region reflective Newton method (MATLAB, The MathWorks, Inc.) to minimize the total $\chi^2$. To quantify the uncertainty in the estimated parameters, we performed 10,000 independent fits with randomly chosen initial parameter sets (within their physiological ranges). This implied the constraint that the spatial order of the components had to be maintained, i.e., we did not allow permutation of the individual groups. In Figs.~S1~and~S2 of the {\it Supplementary Material}, the final $\chi^2$ values are plotted against the final parameters. We followed Ref.~\cite{Fritz_JMB_09} to compute the errors with respect to the optimal parameter values listed in Table~\ref{TABbilayerParameters}. The squared error for parameter $\theta_k$ was calculated using the following equation:
\begin{eqnarray}\label{EQNerrors}
\sigma_{k}^2 & = & \frac{\displaystyle \sum_{\theta_{k,i}} (\theta_{k,i} - \theta_k^{opt})^2 e^{-\chi_i^2/2}}{ \displaystyle\sum_{\theta_{k,i}} e^{-\chi_i^2/2}}\,,
\end{eqnarray}
where $\theta_{k,i}$ is the value of parameter $\theta_k$ in the $i_{th}$ fit, $\theta_k^{opt}$ is the value of $\theta_k$ in the fit with the lowest value of $\chi^2$, and $\chi_i^2$ is the value of $\chi^2$ for the $i_{th}$ fit. In using the likelihood function $e^{-\chi^2/2}$, we assume that the errors in the measurements are independent and normally distributed with widths equal to the standard error of the mean.

\paragraph{\footnotesize Calculation of Anx A2t coverage.} The coverage of Anx A2t was derived from the electron density adjacent to the bilayer as follows: The electron density of the total protein layer $\rho$  can be written as a weighted sum the electron densities of pure protein $\rho_{\mathrm prot}$ and pure water $\rho_{\mathrm{H_2O}}$, i.e.,  $\mathrm{\rho=x \rho_{\mathrm prot}+(1-x) \rho_{\mathrm H_2O}}$, where $x$ and $(1-x)$ are the volume fractions of the protein and water, respectively. As a proxy for the total protein layer we used the maximal electron density in the Anx A2 signature and correspondingly determined the electron density of the pure protein from the chemical sum formula and the mass density of Anx A2. The mass density of Anx A2 is given by $\rho_{m}=\left[1.41+0.145 \exp{(-M/13)}\right]\, g/cm^3$, as described in \citep{fischer2004}. Here $M$ is the molecular mass of Anx A2 in $kDa$, leading to a mass density of $\rho_{m}=1.418\,g/cm^3$. With this, the electron density of Anx A2 is obtained as $\rho_{prot} = 0.458\,e^-/\AA^{3}$ while the electron density of bulk water is given by $\rho_{H_2O} = 0.336\,e^-/\AA^{3}$. Taken together, we find a protein coverage of 91\% in the case of the 50\%\,POPS containing bilayer and 27\% in the case of the 25\%\,POPS containing bilayer.

\normalsize

\section*{Results}

\subsection*{X-Ray Reflectivity}
\begin{figure}
\centerline{\includegraphics[width=7in]{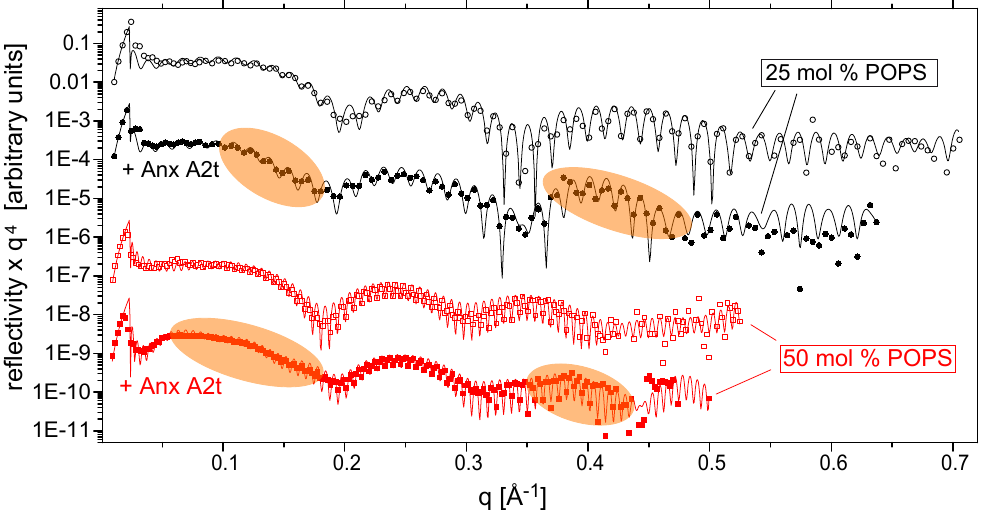}}
\caption{\label{FIGreflectivity} Fritz et al.}
\end{figure}

Anx A2t binds to anionic phospholipids in a $\mathrm{Ca^{2+}}$-dependent manner. In order to scrutinize the structural and dynamical properties of Anx A2t binding, SLBs composed of different ratios of anionic (POPS) to neutral (POPC) lipids were prepared on silicon substrates in Tris 1mM $\mathrm{CaCl_2}$ buffer (calcium buffer hereafter).  For x-ray reflectivity measurements lipid bilayers were mixed in the molar ratios (POPS:POPC) of (1:3), referred to as 25 mol \% POPS and (1:1), referred to as 50 mol \% POPS.  Figure \ref{FIGreflectivity} shows the x-ray reflectivity as a function of the momentum transfer $q$ for both lipid mixtures before and after Anx A2t incubation. The increase in intensity for $q < 0.02\, \mathrm{\AA}^{-1}$ is due to total reflection at the silicon surface and a $q^4$ correction, as described in {\it Materials \& Methods}. The rapid intensity oscillations (Kiessig fringes) show a periodicity of $\Delta q \approx 0.016\, \mathrm{\AA}^{-1}$ for the 25 mol \% POPS containing bilayer and $\Delta q \approx 0.0064\, \mathrm{\AA}^{-1}$ for the 50 mol \% POPS containing bilayer. These narrow fringes stem from the interference between reflections at the silicon oxide layer and the silicon substrate. In contrast, the broad Kiessig fringes with $\Delta q \approx 0.15\, \mathrm{\AA}^{-1}$ have been shown to result from the SLB \cite{Reich_RevSciInstrum_05}. After incubation with Anx A2t, a clear change in signal is observed for both POPS concentrations (compare {\it ellipsoids} in Fig.~\ref{FIGreflectivity}). Importantly, the signal change after Anx A2t incubation is reversible, {\it i.e.}, after rinsing with a calcium gelating buffer (cf. {\it Materials \& Methods}) the A2t-specific signature disappears (data not shown).

\begin{figure}
\centerline{\includegraphics[width=6in]{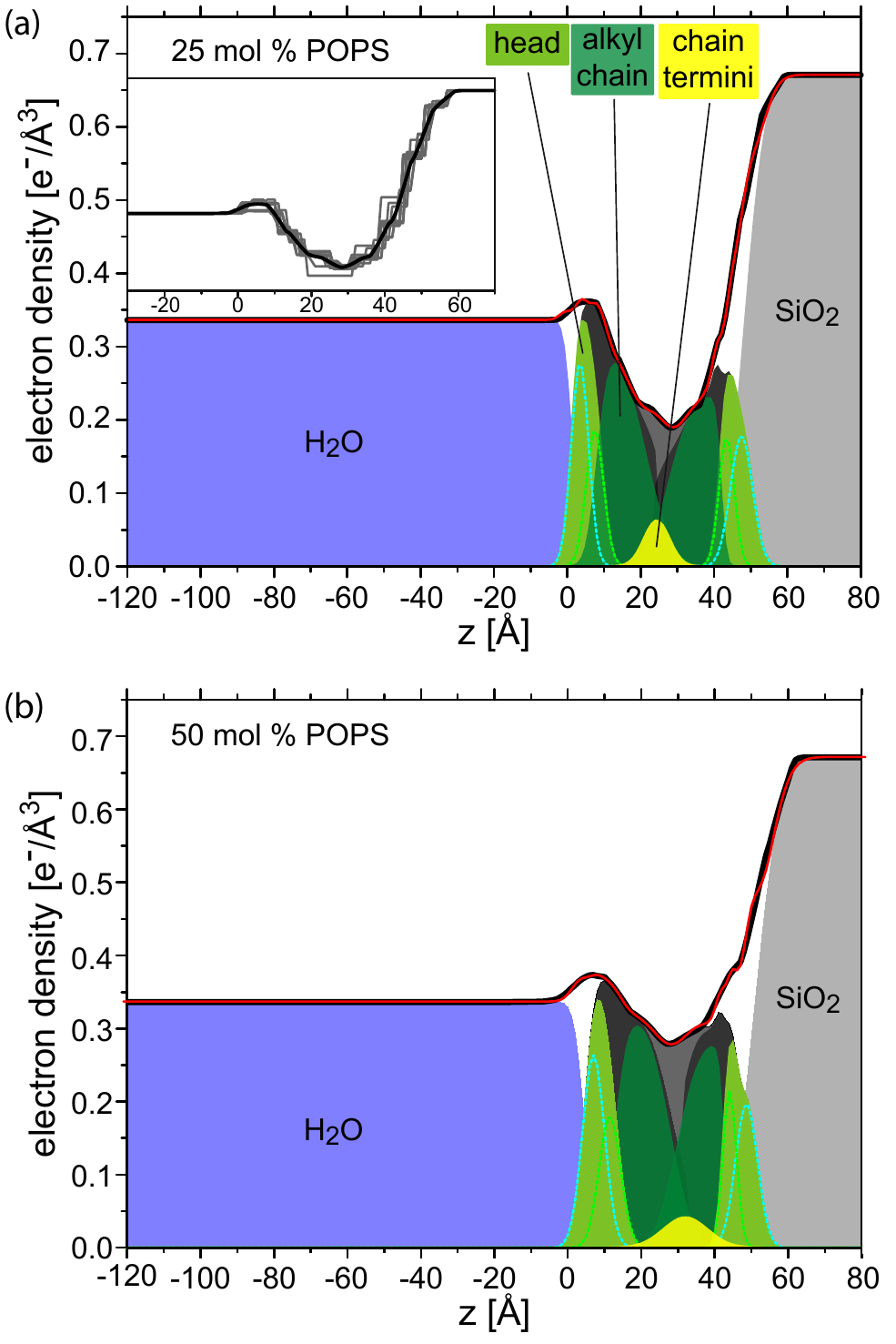}}
\caption{\label{FIGelectronDensity} Fritz et al.}
\end{figure}

To extract structural information of the SLBs before and after protein binding, the x-ray reflectivity data in Fig.~\ref{FIGreflectivity} were analyzed with the help of Parrats algorithm \cite{Parratt_PhysRev_54, Nevot_RevPhysAppl_80}. Briefly, the algorithm estimates the electron density profile that most likely explains the reflectivity data, {\it i.e.}, it yields a fit for the electron density of the probe as a function of the distance from the silicon surface. For each data set in Fig.~\ref{FIGreflectivity} the algorithm was initialized with varying starting parameters and the 15 best fits (lowest $\chi^2$) were selected for arithmetic averaging of the obtained electron density profiles, see exemplarily the inset in Fig.~\ref{FIGelectronDensity}~{\it a} for the SLB containing 25 mol \% POPS. For all details of the fitting procedure please refer to {\it Materials \& Methods}.  The solid lines in Fig.~\ref{FIGreflectivity} show the simulated reflected intensities based on the best fit for each data set and indicate good agreement between experimental data and the quantitative fit. The electron density profiles of the membranes with 25 mol \% POPS (Fig.~\ref{FIGelectronDensity}~{\it a}) and 50 mol \% POPS (Fig.~\ref{FIGelectronDensity}~{\it b}) show both the typical shape of a supported bilayer \cite{Reich_RevSciInstrum_05,Klauda_BiophysJ_06,Kucerka_CurrOpinColloidInterfaceSci_07}: The head distal to the silicon surface as well as the hydrophobic part are clearly visible as an increase and decrease of electron density compared to the buffer's electron density. The head proximal to the silicon substrate is not clearly visible in the total profile of the 25 mol \% POPS containing bilayer (Fig.~\ref{FIGelectronDensity}~{\it a}), however, it is more pronounced for the 50 mol \% POPS containing bilayer (cf. Fig.~\ref{FIGelectronDensity}~{\it b}). The electron density profiles after Anx A2t incubation are shown in Fig.~\ref{FIGelectronDensity_w_prot}. An increase in electron density adjacent to the distal headgroup is observed in both the 25 mol \% POPS and the 50 mol \% POPS containing bilayer. We interpret this increase as an Anx A2t layer. In the case of the 50 mol \% POPS containing SLB the Anx A2t signature is pronounced more strongly compared to the 25 mol \% POPS bilayer, indicating an enhanced  Anx A2t coverage for increasing POPS content. Indeed, as detailed in {\it Materials and Methods}, the 50 mol \% POPS containing bilayer shows 91\% coverage of Anx A2t as compared to 27\% in the case of the 25 mol \% POPS bilayer. Interestingly, both data sets show approximately the same protein layer thickness of about $60\,\AA$, indicating that the higher protein coverage does not influence the Anx A2t configuration. Moreover, already from visual inspection of the electron density profiles the vertical configuration of Anx A2t seems very unlikely, as it would result in a significantly larger thickness of the protein layer \cite{Nakata_JCellBiol_90,Sopkova-deOliveiraSantos_BiochimBiophysActa_00,Lambert_JBiolChem_04,Schulz_Proteins_07}. A more quantitative analysis of the electron density profiles follows below. 

\subsection*{Decomposition of Electron Density Profiles}
To extract quantitative information about the spatial dimensions of the protein layer and to study possible conformational changes in the SLB, the electron density profiles were analyzed in more detail. To this end, the electron density profiles were divided into several component groups of the lipid bilayer and the Anx A2t complex (Table~\ref{TABinputParameters}). These groups contribute additively to the total electron density: two Gaussian functions were used for the headgroup (phosphate + $\mathrm{Ca}^{2+}$ + serine/choline) and the backbone (glycerol + carbonyl group), two error functions were used for the alkyl chains and one Gaussian function for the chain termini (methyl groups) of both leaflets \cite{Kucerka_BiophysJ_08}.  The areas of all groups were fixed to the stoichiometric ratios of their electron numbers, cf. Table~\ref{TABinputParameters}, such that only the total area of each leaflet, the width and the position of each group were fitting parameters\footnote{The area of the gaussian for the methyl groups was adjusted in stoichiometric proportion to the arithmetic mean of the areas of the alkyl chains in both leaflets.}. In addition, the area per lipid of each leaflet was constrained to values above $40\, \mathrm{\AA}^2$/lipid, since film balance measurements of PS lipids in calcium buffer revealed this value as an empirical minimum for their area fraction \cite{Mattai_Biochemistry_89}. Silicon substrate and water are each represented by an error function. The silicon substrate's roughness was adjusted to the value of $4\,\AA$ obtained by independent measurements of the bare wafers and was kept constant during the fitting process.  Anx A2t is represented by 6 error functions, where 2 error functions represent the p11 dimer and 4 error functions account for the hydrophilic and hydrophobic domains of the Annexin monomer, respectively. The thickness of the protein layer was calculated by the half-maximal width of the two enveloping error functions. The separation of the total electron density into component groups is not obvious per se, however, the physical constraint that the total contribution of each lipid group was fixed to its electron number (see Table~1, \cite{Feigenson_Biochemistry_86,Mattai_Biochemistry_89,Klauda_BiophysJ_06}) results in four unambiguous and stable fitting parameters, namely the (inverse) packing density of the lipids, {\it i.e.}, the area per lipid in each leaflet, the head-to-head distance and the width of the protein layer. These parameters before and after Anx A2t binding are summarized in Table~2 and will be discussed further below.

\begin{figure}
\centerline{\includegraphics[width=6in]{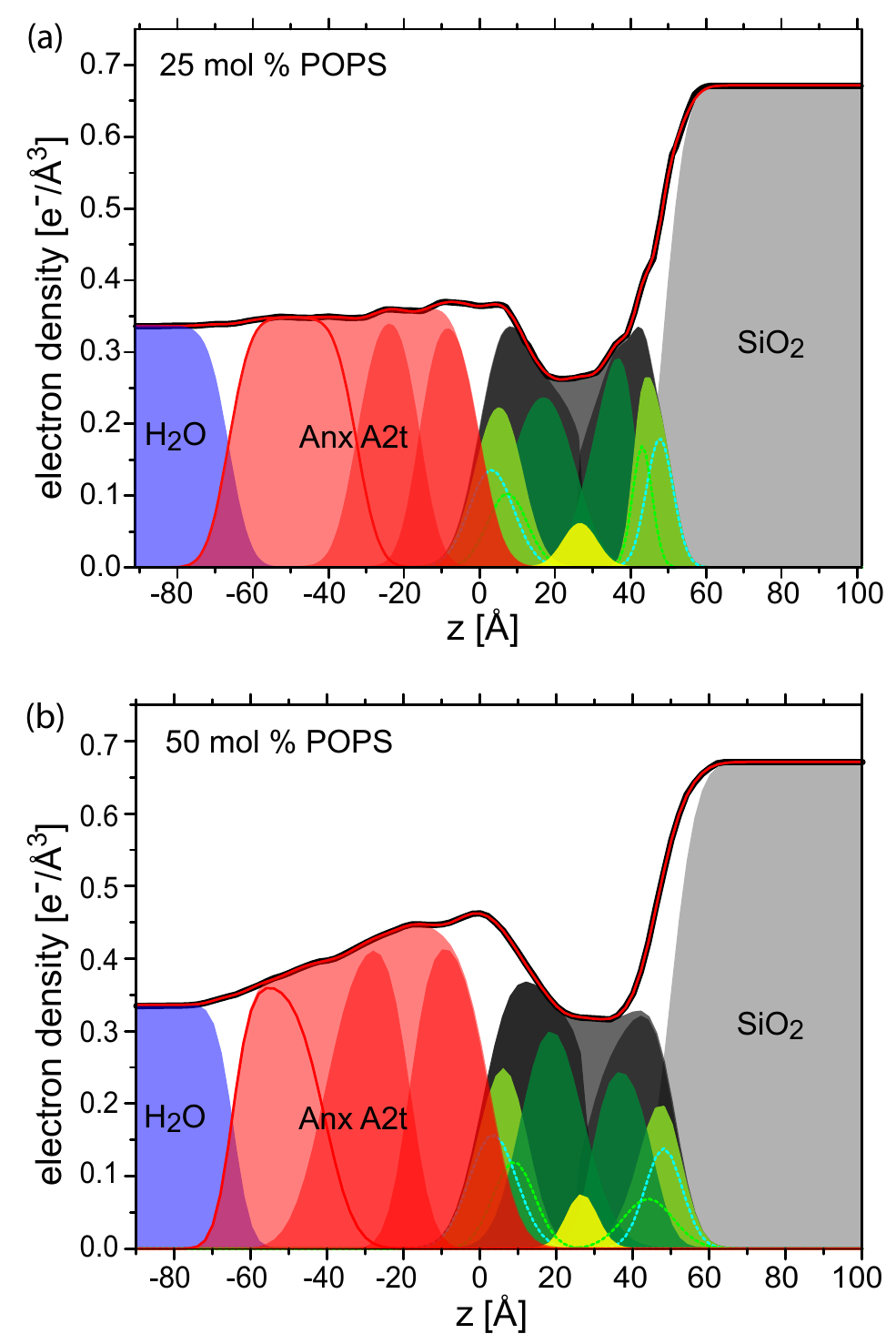}}
\caption{\label{FIGelectronDensity_w_prot} Fritz et al.}
\end{figure}

\paragraph*{Conformation of the Anx A2t Complex.}
The spatial decomposition of the overall electron density profile into the different groups is shown in Figs.~\ref{FIGelectronDensity}~and~\ref{FIGelectronDensity_w_prot} (see figure caption for color code).
The thickness of the resulting Anx A2t layer is $59 \pm 6\, \mathrm{\AA}$  for the 25 mol \% and $67\pm5\, \mathrm{\AA}$ for the 50 mol \% POPS containing bilayer (see {\it red areas} in Figs.~\ref{FIGelectronDensity_w_prot}~{\it a}~and~{\it b}). This width is  remarkably close to the dimensions of the side-by-side configuration of Anx A2t ($56\, \mathrm{\AA}\, \mathrm{to}\, 61\, \mathrm{\AA}$ \cite{Schulz_Proteins_07}), cf. Fig.~\ref{FIGscheme}~({\it a, top}). In contrast, the vertical arrangement depicted in Fig.~\ref{FIGscheme}~({\it a, bottom}) is expected to result in an Anx A2t thickness between $90\, \mathrm{\AA}\, \mathrm{and}\, 107 \, \mathrm{\AA}$ \cite{Lambert_JMolBiol_97,Nakata_JCellBiol_90} and can thus be ruled out by our data. Hence, our reflectometry experiments favor the side-by-side configuration as the most plausible mechanism for Anx A2t binding to single SLBs {\it in vitro}.

\paragraph*{Structural Changes in the SLB.}
Next, we analyzed the structural reorganization of the lipid bilayer associated with changes in the lipid ratio and with binding of Anx A2t. To this end, we used the head-to-head distance of the lipids as a proxy for the thickness and the area per lipid as a measure for the (inverse) packing density of the SLB, see Table~\ref{TABbilayerParameters}. In the presence of calcium-containing buffer, the head-to-head distance is $44\pm 1\, \mathrm{\AA}$ for the 25 mol \% POPS bilayer and $42\pm1\,  \mathrm{\AA}$ for the 50 mol \% POPS bilayer. After Anx A2t incubation the SLB thickness increases slightly to $48\pm3\,  \mathrm{\AA}$ for the 25 mol \% POPS bilayer and to $45\pm 3\,  \mathrm{\AA}$ for the 50\% POPS bilayer, which is in both cases within the resolution tolerance of the experiment. That is, the head to head distance does not vary significantly between the samples. In contrast, the packing density displays a much stronger response to both the fraction of anionic lipid and to binding of Anx A2t: First of all, we observed that all anionic bilayers exhibit an asymmetric packing density of the two leaflets, {\it i.e.}, the distal (Annexin-facing) leaflet contains more lipids per unit area than the proximal (surface-facing) leaflet. In detail, for the 25 mol~\% POPS sample in calcium-containing buffer the area per lipid is $60\pm2\,\mathrm{\AA}^2$/lipid for the distal leaflet compared to  $69\pm 3\,\mathrm{\AA}^2$/lipid for the proximal leaflet. An increase of the POPS fraction in the lipid mixture leads to a densification of the bilayer and increases its asymmetry; in the 50 mol \% POPS sample the area fraction is $51\pm3 \,\mathrm{\AA}^2$/lipid for the distal as compared to $67\pm 2 \,\mathrm{\AA}^2$/lipid for the proximal leaflet. The binding of Anx A2t to the bilayer potentiates the asymmetric densification of the two leaflets: For the 25 mol \% POPS sample the area fraction decreases to $46\pm 3 \,\mathrm{\AA}^2$/lipid in the distal compared to $65\pm 5\,\mathrm{\AA}^2$/lipid in the proximal leaflet. For the 50 mol \% POPS sample we obtain $41\pm 4 \,\mathrm{\AA}^2$/lipid in the distal as compared to $59\pm 4 \,\mathrm{\AA}^2$/lipid for the proximal leaflet.

Note that in our decomposition procedure we explicitly allowed for the hypothetical scenario in which Anx A2t penetrates the headgroup of the bilayer to a certain degree. On the first sight it may appear as if this could be an alternative explanation for the increased electron density in the buffer-facing leaflet.  However, it turned out that this hypothetical scenario did not reproduce the electron density profiles appropriately. This is because the stoichiometric ratio between the lipid components is fixed during our fitting process. Specifically, if Anx A2t would penetrate the headgroup of the bilayer more deeply, not only the electron density of the headgroups, but also the density of the lipid chains would be reduced. Apparently, this reduction of the density in the central region of the bilayer cannot be compensated by an even deeper penetration of the Anx A2t in our fitting procedure. Therefore we believe that the increased electron density in the buffer-facing leaflet is indeed due to a densification of the lipids.

\subsection*{Continuous Photobleaching}
\begin{figure}
\centerline{\includegraphics[width=6in]{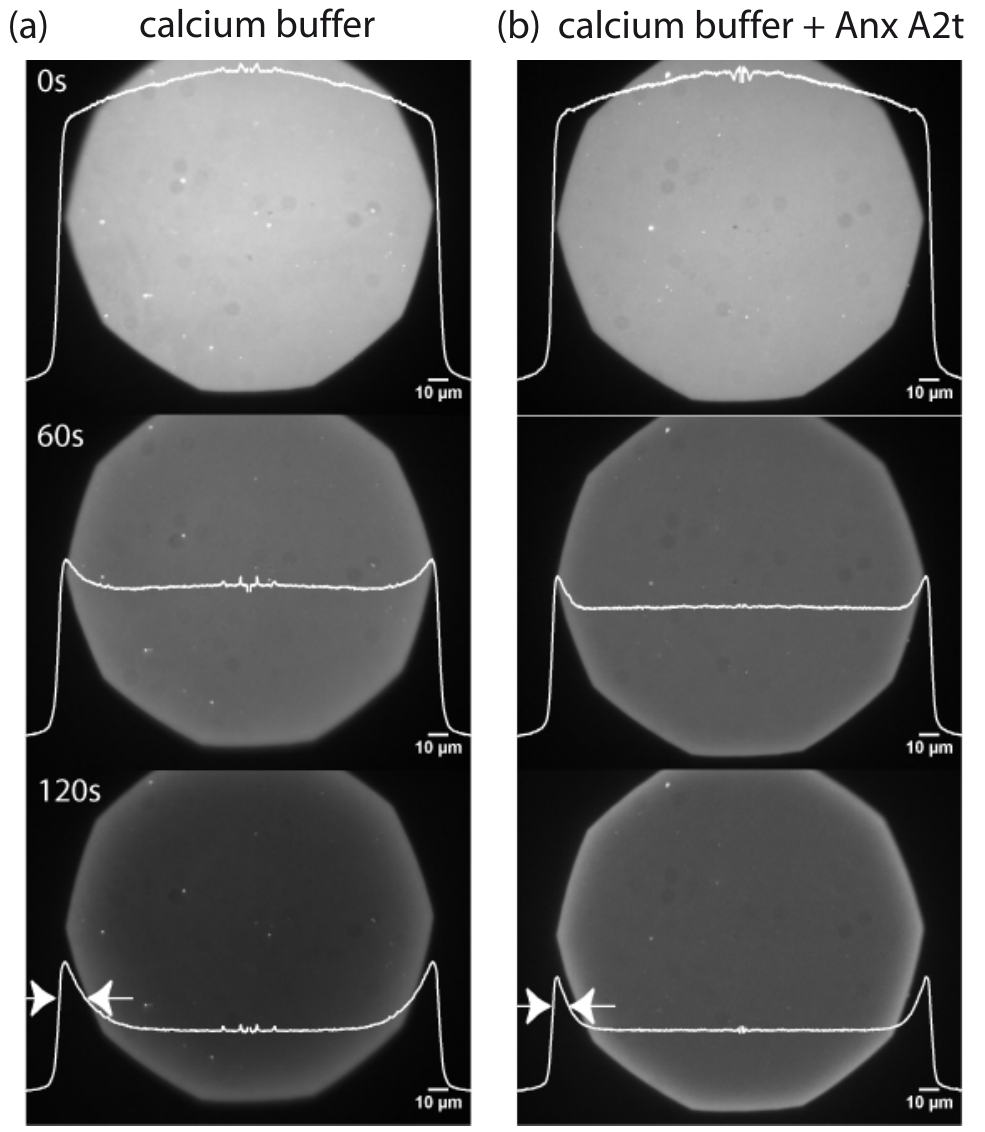}}
\caption{\label{FIGbleaching} Fritz et al. }
\end{figure}

One might expect that the structural changes of the SLB in response to the anionic fraction of lipids and the binding of Anx A2t should also induce changes in the dynamical properties of the bilayer. In particular, the diffusion properties of the lipids should vary strongly with bilayer density. Therefore, we used continuous photobleaching to test whether the densification of the SLBs is correlated with a reduction in lipid mobility. To this end, a small fraction of fluorescently labeled lipids was added to the lipid mixture and a circular spot was continuously illuminated by the lamp of the microscope, see {\it Materials \& Methods}. The interplay between continuous photobleaching of the dyes inside the spot and diffusion of unbleached molecules into the spot leads to a characteristic, time dependent intensity profile; the higher the diffusion constant, the broader the profile at the rim of the spot. Similar to our XR measurements, we studied SLBs with 50 mol \% POPS and 25 mol \% POPS, as well as a pure POPC bilayer as a control. A time series of a 25 mol \% POPS containing bilayer during continuous bleaching before and after protein incubation is shown in Figs.~\ref{FIGbleaching}~{\it a} and {\it b}, respectively. The white curves are the radially averaged intensity profiles of each exposure. Both time series show the same bleaching rate in the central region. At the rim of the field-of-view, a diffusion-induced increase of the fluorescence is observed. After incubation with Anx A2t, the fluorescence profile at the rim narrows and weakens (cf. {\it arrows} in Fig.~\ref{FIGbleaching}), indicating a lower diffusion constant of the lipids after Anx A2t binding.

\begin{figure}
\centerline{\includegraphics[width=6in]{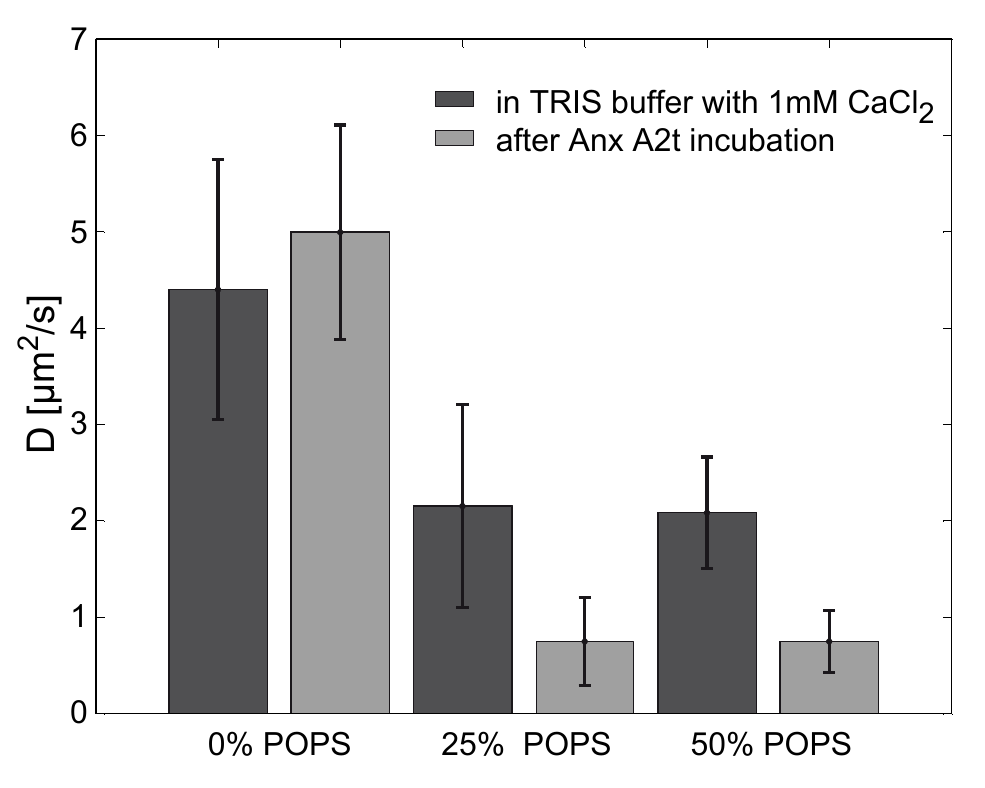}}
\caption{\label{FIGdiffusionConstants} Fritz et al.}
\end{figure}

Quantitative information of the diffusion constant of the labeled lipids can be obtained from the evaluation of such profiles \cite{Dietrich_BiophysJ_97,Hochrein_Langmuir_06}. Briefly, a timeseries of the radially averaged intensity profiles is extracted from the fluorescence images and the bleaching rate is estimated in the center of the bleached area. The intensity profiles were corrected for uneven illumination and the radial reaction-diffusion equation, describing bleaching and lipid mobility, is solved numerically. Subsequently, the diffusion constant is used as the sole fit-parameter to minimize the $\chi^2$ between experimental and theoretical timeseries of bleaching profiles. In Fig.~\ref{FIGdiffusionConstants} we show the resulting diffusion constants of the three different bilayer compositions in calcium buffer before ({\it dark grey bars}) and after Anx A2t incubation ({\it light grey bars}). For the pure POPC bilayer we obtain a diffusion constant of $D = 4.2 \pm 1.4\, \mu m^2/s$ before and $D = 5.0 \pm 1.2\, \mu m^2/s$ after Anx A2t incubation, respectively. This result indicates that the supported lipid bilayer is fluid and displays a diffusion constant within the range observed in previous studies \cite{Ladha_BiophysJ_96,Liangfang_JChemPhys_05,Scomparin_EurPhysJ_09}. It also reveals that the mobility of the lipids in the pure POPC bilayer is not influenced by incubation with Anx A2t, as expected from the fact that the negatively charged POPS is essential for $\mathrm{Ca^{2+}}$-dependent Annexin binding. In calcium-containing buffer the 25 mol \% and 50 mol \% POPS containing bilayer show similar diffusion constants of $D = 2.3 \pm 1.0 \, \mu m^2/s$ and $D = 2.0 \pm 0.8\, \mu m^2/s$, respectively. After Anx A2t incubation, the diffusion constant of the two POPS containing SLBs is significantly reduced to $D = 0.9 \pm 0.5\,  \mu m^2/s$ and $0.7 \pm 0.2\, \mu m^2/s$. Hence, both anionic SLBs show the same reduction in diffusion constant within the error bar. Note that we did not measure the reference value without calcium buffer, since the 50 mol\,\% POPS bilayers tend to delaminate from the silicon substrate without the stabilizing effect of calcium buffer.

\section*{Discussion}

 In this paper we investigated the conformation of the Anx A2t complex upon binding to single surface supported bilayers and the accompanied structural and dynamical changes in the lipid membranes. The thicknesses of the Anx A2t layer obtained from our x-ray measurements indicate that Anx A2t unlikely binds in the vertical conformation to a single membrane, as it would result in a  significantly larger protein thickness \cite{Lambert_JMolBiol_97,Nakata_JCellBiol_90}. Instead, our results  favor the side-by-side configuration of the Anx A2 tetramer \cite{Schulz_Proteins_07} and thus, provide a non-invasive and independent verification of previous AFM studies \cite{Menke_ChemBioChem_04}, see Fig.~\ref{FIGsummary} for a summarizing illustration.

Binding of Anx A2t to single membranes in a side-by-side configuration has potential implications for the route of Anx A2t-induced membrane bridging. On the one hand it seems plausible that tetramers in the side-by-side configuration perform some kind of ``breathing modes'', in which one of the two Annexin monomers  temporarily detaches from its membrane interface  and is free to bind to an approaching bilayer. As a result, the membrane-membrane contact would be established by Anx A2t in a vertical configuration, cf. Fig.~\ref{FIGscheme}~({\it b, bottom}), in line with cryo-electron microscopy results on Anx A2t-connecting vesicles \cite{Lambert_JMolBiol_97}. However, this mode of membrane-bridging demands a high flexibility of the Anx A2t complex. Although previous studies indicated that the p11 dimer displays a certain flexibility  \cite{Rety_NatStructBiol_99}, it is currently unknown whether the Anx A2t tetramer is indeed able to fluctuate between vertical and side-by-side configuration. Alternatively, it was suggested that the membrane-membrane contact is mediated by the formation of a heterooctameric structure composed of two opposing Anx A2t complexes, cf. Fig.~\ref{FIGscheme}~({\it b, top}). This molecular arrangement was favored by Waisman in the case of Anx A2t-chromaffin granules interactions \cite{WaismanM_MolCellBiochem_95} and has also been discussed to occur following disulfide-bridge formation between cysteines within the C-terminal region of p11 \cite{Lewit-Bentley_CellBiolInt_00}. In this model the dynamics of octamer-formation could take two alternative routes: Either the octamers pre-assemble on a single membrane interface and thus directly allow bridging to a second membrane, or the Anx A2t tetramers distribute among both membrane interfaces and form octamers only upon membrane-membrane contact. From our x-ray experiments we found that the thickness of the protein layer on a single SLB is only compatible with a monolayer of Anx A2t, suggesting that a pre-assembly of octamers on a single membrane interface is not significant under the experimental conditions used here.

\paragraph{Structural changes in the lipid bilayer.}
Typical densities of uncharged bilayers are around $70\,\mathrm{\AA}^2$/lipid \cite{Vacklin_Langmuir_05,Klauda_BiophysJ_06}. Indeed, for the silicon-facing leaflet we found  packing densities of $69\pm 3 \,\mathrm{\AA}^2$/lipid for the 25 mol\% POPS  and $67\pm 2\,\mathrm{\AA}^2$/lipid for the 50 mol\% POPS containing bilayer. Hence, the density of the substrate-facing leaflet turns out to be independent on POPS amount and is close to values known for uncharged membranes. In contrast, in the presence of $\mathrm{Ca^{2+}}$ the distal leaflet density is increased to $60\pm 2 \,\mathrm{\AA}^2$/lipid for the 25 mol \% POPS and to $51\pm 3\,\mathrm{\AA}^2$/lipid for the 50 mol \% POPS containing bilayer. Thus, in all measurements, the distal, buffer-facing leaflet shows a pronounced and POPS dependent response to the presence of calcium-containing buffer. Indeed, it is known that for mixtures of anionic and zwitterionic lipids, $\mathrm{Ca^{2+}}$ induces molecular segregation and clustering, and the formation of domains with a higher density \cite{Trauble_PNAS_74}.  Based on these findings, we propose that the response of the buffer-facing leaflet to $\mathrm{Ca^{2+}}$ results from a chelating effect of calcium ions. Indeed, a chelating effect of calcium ions bridging anionic lipids has been reported before \cite{Casal_Biochemistry_87,Mattai_Biochemistry_89}: In the distal leaflet $\mathrm{Ca^{2+}}$ bridges at least two POPS molecules and thus results in a closer packing density, while in the proximal leaflet, if POPS is not depleted by electrostatics (see below), $\mathrm{Ca^{2+}}$ rather bridges each individual POPS molecule to the negatively charged silicon substrate, leaving the density unchanged.

Our data was obtained at a salt concentration of 20 mM Tris/HCl, 100 mM NaCl, 1 mM $\mathrm{CaCl_2}$. Here, the Debye screening length is about $9\,\mathrm{\AA}$ only, and the question emerges whether electrostatic repulsion may also favor an enrichment of the anionic lipid in the distal leaflet, thus contributing to the stronger response of this leaflet to $\mathrm{Ca^{2+}}$. On silicon supports, anionic Texas Red DHPE lipids in a POPC matrix have been found to be enriched in the distal leaflet at moderate (75 mM) monovalent salt concentrations \cite{Shreve_Langmuir_08}. This observation was accounted to repulsion from anionic hydroxyl groups at the silicon surface screened by the associated Debye length of approximately $11\,\mathrm{\AA}$. Since the water gap between silicon support and surface supported bilayers has been shown to be below detection limit for reflectivity studies \cite{Reich_RevSciInstrum_05, Horton_Langmuir_07, Reich_BiophysJ_08}, it is well below the Debye screening length. In this context one should also note that screening within the hydrophobic core of the SLB is expected to be low and thus, if the proximal leaflet is not screened sufficiently, also the distal leaflet will be under the influence of the surface electric field \cite{Shreve_Langmuir_08}. Therefore a charge-induced POPS enrichment in the distal leaflet may also contribute to the investigated asymmetric densification. It has been pointed out that the enrichment mechanism itself may occur at the early stage of vesicle spreading by rapid diffusion of lipids between the two leaflets via edge effects rather than via flip-flop \cite{Richter_Langmuir_04}. Here we used the method of vesicle spreading with the help of osmotic pressure. Vesicle spreading was performed in pure water and thus a redistribution of negatively charged lipid during vesicle spreading is very likely. Note that for $\mathrm{SiO_2}$ the density of hydroxyl groups and thus the negative charge at the surface depends crucially on the cleaning procedure. Here we tried to maximize the amount of hydroxyl groups by aggressive wet-chemical cleaning, see {\it Materials \& Methods}. Less efficient cleaning procedures or longer storage times after cleaning may result in a reduced number of hydroxyl groups, which may be the origin for some controversy in the literature \cite{Richter_Langmuir_04, Shreve_Langmuir_08}.

Upon binding of Anx A2t, for the 25\% POPS containing bilayer the density of the distal leaflet increased significantly, whereas the density of the proximal leaflet was not notably affected. In the 50 \% POPS containing bilayer both leaflets show a further, albeit still asymmetric, densification. Interestingly, Langmuir compression experiments with DMPA suggested a phase transition between liquid expanded to liquid condensed phase at a packing density of $40 \,\mathrm{\AA}^2$/lipid  \cite{Schalke_BiochimBiophysActa_00}, similar to the densities we found for the buffer-facing leaflet. Likewise, Watkins et al. have reported a comparable packing density for DPPC SLBs \cite{Watkins_PhysRevLett_09}, in agreement with gel phase data. Thus, it appears as if the responses of both leaflets are decoupled unless the density in the distal leaflet reaches gel phase values. Only then also the proximal leaflet "senses" the binding Anx A2t and gets compactified. Hence, it is tempting to speculate that the Anx A2t-mediated densification of the bilayer is accompanied with a transition to gel phase in the distal leaflet. Indeed, asymmetric phase transitions in only a single leaflet of the bilayer have been observed before, {\it e.g.}, in dilauroylphosphatidylcholine/distearoylphosphatidylcholine (DLPC/DSPC) mixtures \cite{Lin_BiophysJ_06}.

\paragraph{Dynamical changes in the bilayer.} Continuous bleaching measurements show a reduced diffusion constant of $D = 2.3 \pm 1.0\,\mu m^2/s$ for POPS containing SLBs in calcium containing buffer compared to $D = 4.2 \pm 1.4\,\mu m^2/s$ for pure POPC SLBs. This decrease in diffusion constant could be accounted to several mechanisms. The first mechanism is based on obstructed diffusion, in which calcium ions induce lipid domains of higher packing density and thereby form obstacles for the diffusion of fluorescently labeled lipids \cite{Ratto_BiophysJ_02,Fenz_Langmuir_09}. Alternatively, since $\mathrm{Ca^{2+}}$ promotes the spreading of negatively charged small unilamellar vesicles on $\mathrm{SiO_2}$ \cite{Cremer_JPhysChemB_99,Richter_Langmuir_06}, one may attribute the reduced diffusion constant to partial sticking \cite{Dertinger_Langmuir_06} of POPS to some ion-bridged OH groups. Our x-ray results indicate that most of the structural changes due to $\mathrm{Ca^{2+}}$ are confined to the buffer-facing leaflet, suggesting that the formation of domains could indeed be the origin of the reduced lipid mobility. Yet, with our experiments we cannot exclude that sticking of POPS to the surface also plays a role. However, since only small amounts of lipids are absorbed to the surface at any time \cite{Dertinger_Langmuir_06}, one may speculate that obstructed diffusion is the main mechanism for the reduction of lipid mobility. Obstructed diffusion in phase-separated SLBs has been studied before, and a reduction of the diffusivity by 50 \% was observed for an area fraction of the gel-phase of 0.4 \cite{Ratto_BiophysJ_02}. Hence, due to a higher lipid density in the 50\% POPS containing bilayer we expected that a higher POPS content leads to a lower diffusion constant. However, we could not resolve this reduction within experimental error. Interestingly, Gilmanshin {\it et. al} have observed a similar behavior for mixtures of POPC and anionic POPG \cite{gilmanshin1994}: The diffusion constant was not dependent on POPC concentration between values of 0 and 80 mol\,\% in calcium containing buffer, whereas it slightly rose with higher POPC amount. Thus, our measurements show similar behavior for mixtures of POPS and POPC.

Upon Anx A2t incubation, the diffusion constant of the two POPS containing SLBs is significantly reduced to $0.9 \pm 0.5 \,\mu m^2/s$ (25 mol \% POPS) and $0.7 \pm 0.2 \,\mu m^2/s$ (50 mol\% POPS). Hence, both anionic bilayers show the same reduction within the error bars, while the POPC control sample displays no decrease in mobility upon protein binding. FRAP measurements with surface supported POPC/POPG membranes also revealed a decrease in diffusion constant upon binding of Annexin IV \cite{gilmanshin1994} and the diffusion constant decreased with increasing POPG fraction. However, above a threshold of 50 mol \% POPG no further reduction of the mobility was observed. From our measurements it seems as if in POPC/POPS mixtures after Anx A2t binding this threshold appears at slightly lower amount of anionic POPS. Yet, from our data it cannot be excluded that there is still a modest dependency of the diffusion constant within the experimental errors.

Lipid demixing upon protein binding to multicomponent membranes has been discussed before in the context of oppositely charged lipid-protein pairs \cite{Heimburg_BiophysJ_99,May_BiophysJ_00}.  Menke et al. \cite{Menke_ChemBioChem_04} have shown by AFM measurements that an area of POPS enrichment develops around Anx A2t. This suggests that obstructed diffusion by phase separation could also be the origin for the reduction of diffusion in presence of Anx A2t, presumably due to POPS assembly below Anx A2t \cite{Dermine_JBiolChem_01}. It was speculated that Anx A2t may act directly to trap and cluster PS, thereby creating microdomains in the plasma membrane \cite{Chasserot2005}. Our x-ray data supports this picture: In all measurements the bilayers show a higher density after Anx A2t binding.

\section*{Conclusions}
Our observations might have important consequences for the understanding of the physiological processes induced by membrane-associated Anx A2t. It was proposed that domain formation may act as nucleation site for lipid rafts and promote their clustering. Once domains are formed, raft structures and the associated cholesterol may further stabilize the lipid-Annexin 2 interaction {\it in vivo}, resulting in Annexin 2-membrane scaffolds that are required to assemble components of the exocytotic machinery \cite{Chasserot2005}. However, the role of Anx A2t in this mechanism remained unclear. In addition, it was proposed that Anx A2t creates pores through a densification of the bilayer and thereby facilitates membrane fusion \cite{faure2002}.  Here we could for the first time demonstrate that the protein is indeed able to induce a densification of PS-containing bilayers {\it in vitro} and, in addition, resolve that  primarily the protein-facing leaflet is compactified. It is tempting to speculate that this asymmetry might be involved in the mechanism of Anx A2t-mediated endo- and exocytosis: The asymmetric insertion of lipids into the outer monolayer of lipid vesicles is often accompanied by positive-curvature strain \cite{Esteban-Martin2009}. In fact, Monte Carlo simulations showed that phase separation in asymmetric bilayers leads to spontaneous budding of the membrane \cite{Wallace2005}. In the future it will be interesting to use off-specular neutron scattering techniques for the study of Anx A2t-induced membrane reorganization. For instance, deuterated POPS could be leveraged to enhance the contrast with respect to POPC and thus, to monitor protein-induced lipid segregation and clustering in the two leaflets.


\section*{Acknowledgements}
Financial support was obtained from BMBF (05KN7WMA) and from the DFG Nanosystems-Initiative-Munich (NIM). Support from Wolfgang Caliebe, Oliver Seeck (beamline D4) and Oier Bikondoa (beamline ID01) as well as travel support from HASYLAB/DESY (Hamburg) and ESRF (Grenoble) is gratefully acknowledged.

\clearpage


\begin{thebibliography}{57}
\providecommand{\natexlab}[1]{#1}

\bibitem{Biener_JBiolChem_96}
Biener, Y., R.~Feinstein, M.~Mayak, Y.~Kaburagi, T.~Kadowaki, and Y.~Zick.
  1996.
\newblock {Annexin II is a novel player in insulin signal transduction}.
\newblock \emph{J. Biol. Chem.} 271:29489--29496.

\bibitem{Filipenko_JBiolChem_04}
Filipenko, N., T.~MacLeod, C.-S. Yoon, and D.~Waisman. 2004.
\newblock {Annexin A2 is a novel RNA-binding protein}.
\newblock \emph{J. Biol. Chem.} 279:8723--8731.

\bibitem{Rety_NatStructBiol_99}
R{\'e}ty, S., J.~Sopkova, M.~Renourd, S.~Tabares, D.~Osterloh, V.~Gerke,
  F.~Russo-Marie, and A.~Lewit-Bentley. 1999.
\newblock {A crystal structure of a complex of p11 with the annexin II
  N-terminal peptide}.
\newblock \emph{Nat. Struct. Biol.} 6:89--95.

\bibitem{Burger_JMolBiol_96}
Burger, A., R.~Berendes, S.~Liemann, J.~Benz, A.~Hofmann, P.~G\"ottig,
  R.~Huber, V.~Gerke, C.~Thiel, J.~R\"omisch, and K.~Weber. 1996.
\newblock {The crystal structure and ion channel activity of human annexin II,
  a peripheral membrane protein}.
\newblock \emph{J. Mol. Biol.} 257:839--847.

\bibitem{Rosengarth_Annexins_04}
Rosengarth, A., and H.~Luecke. 2004.
\newblock {Annexin A2 - does it induce membrane aggregation by a new multimeric
  state of the protein?}
\newblock \emph{Annexins}. 1:129--136.

\bibitem{Sopkova-deOliveiraSantos_BiochimBiophysActa_00}
Sopkova-de Oliveira~Santos, J., F.~Oling, S.~R{\'e}ty, A.~Brisson, J.~Smith,
  and A.~Lewit-Bentley. 2000.
\newblock {S100 protein-annexin interactions: a model of the (Anx2-p11)2
  heterotetramer complex}.
\newblock \emph{Biochim. Biophys. Acta}. 1498:181--191.

\bibitem{Menke_ChemBioChem_04}
Menke, M., M.~Ross, V.~Gerke, and C.~Steinem. 2004.
\newblock {The molecular arrangement of membrane-bound annexin A2-S100A10
  tetramer as revealed by scanning force microscopy}.
\newblock \emph{ChemBioChem}. 5:1003--1006.

\bibitem{Lewit-Bentley_CellBiolInt_00}
Lewit-Bentley, A., S.~R{\'e}ty, J.~Sopkova-de Oliveira~Santos, and V.~Gerke.
  2000.
\newblock S100 protein complexes: some insights from structural studies.
\newblock \emph{Cell. Biol. Int.} 24:799--802.

\bibitem{Schulz_Proteins_07}
Schulz, D., S.~Kalkhof, A.~Schmidt, C.~Ihling, C.~Stingl, K.~Mechtler,
  O.~Zsch\"ornig, and A.~Sinz. 2007.
\newblock Annexin a2/p11 interaction: New insights into annexin a2 tetramer
  structure by chemical crosslinking, high-resolution mass spectrometry, and
  computational modeling.
\newblock \emph{Proteins}. 69:254--269.

\bibitem{Lambert_JMolBiol_97}
Lambert, O., V.~Gerke, M.~F. Bader, F.~Porte, and A.~Brisson. 1997.
\newblock Structural analysis of junctions formed between lipid membranes and
  several annexins by cryo-electron microscopy.
\newblock \emph{J. Mol. Biol.} 272:42--55.

\bibitem{Nakata_JCellBiol_90}
Nakata, T., K.~Sobue, and N.~Hirokawa. 1990.
\newblock {Conformational change and localization of calpactin I complex
  involved in exocytosis as revealed by quick-freeze, deep-etch electron
  microscopy}.
\newblock \emph{J. Cell. Biol.} 110:13--25.

\bibitem{Lambert_JBiolChem_04}
Lambert, O., N.~Cavusoglu, J.~Gallay, M.~Vincent, J.~L. Rigaud, J.-P. Henry,
  and J.~Alaya-Sanmartin. 2004.
\newblock Novel organisation and properties of annexin 2-membrane complexes.
\newblock \emph{J. Biol. Chem.} 279:10872--10882.

\bibitem{Miller_PhysRevLett_05}
Miller, C.~E., J.~Majewski, T.~Gog, and T.~L. Kuhl. 2005.
\newblock {Characterization of biological thin films at the solid-liquid
  interface by X-ray reflectivity}.
\newblock \emph{Phys. Rev. Lett.} 94:238104.

\bibitem{Reich_RevSciInstrum_05}
Reich, C., M.~B. Hochrein, B.~Krause, and B.~Nickel. 2005.
\newblock A microfluidic setup for studies of solid-liquid interfaces using
  x-ray reflectivity and fluorescence microscopy.
\newblock \emph{Rev. Sci. Instrum.} 76:095103.

\bibitem{Nickel_Biointerphases_08}
Nickel, B. 2008.
\newblock Nanostructure of supported lipid bilayers in water.
\newblock \emph{Biointerphases}. 3:FC40.

\bibitem{Novakova_PhysRevE_06}
Novakova, E., K.~Giewekemeyer, and T.~Salditt. 2006.
\newblock {Structure of two-component lipid membranes on solid support: An
  x-ray reflectivity study}.
\newblock \emph{Phys. Rev. E}. 74:051911.

\bibitem{Reich_BiophysJ_08}
Reich, C., M.~R. Horton, B.~Krause, A.~P. Gast, J.~O. R\"adler, and B.~Nickel.
  2008.
\newblock Asymmetric structural features in single supported lipid bilayers
  containing cholesterol and g(m1) resolved with synchrotron x-ray
  reflectivity.
\newblock \emph{Biophys. J.} 95:657--668.

\bibitem{Horton_Langmuir_07}
Horton, M.~R., C.~Reich, A.~P. Gast, J.~O. R\"adler, and B.~Nickel. 2007.
\newblock Structure and dynamics of crystalline protein layers bound to
  supported lipid bilayers.
\newblock \emph{Langmuir}. 23:6263--6269.

\bibitem{Chasserot2005}
Chasserot-Golaz, S., N.~Vitale, E.~Umbrecht-Jenck, D.~Knight, V.~Gerke, and
  M.-F. Bader. 2005.
\newblock {Annexin 2 promotes the formation of lipid microdomains required for
  calcium-regulated exocytosis of dense-core vesicles}.
\newblock \emph{Mol. Biol. Cell}. 16:1108--1119.

\bibitem{ross2005}
Ross, M., V.~Gerke, and C.~Steinem. {2003}.
\newblock {Membrane composition affects the reversibility of annexin A2t
  binding to solid supported membranes: A QCM study}.
\newblock \emph{Biochemistry}. {42}:{3131--3141}.

\bibitem{Gerke_EMBOJ_84}
Gerke, V., and K.~Weber. 1984.
\newblock Identity of {p36K} phosphorylated upon rous-sarcoma virus
  transformation with a protein purified from brush-borders - calcium-dependent
  binding to non-erythroid spectrin and {F-actin}.
\newblock \emph{EMBO J.} 3:227--233.

\bibitem{Parratt_PhysRev_54}
Parratt, L.~G. 1954.
\newblock Surface studies of solids by total reflection of x-rays.
\newblock \emph{Phys. Rev.} 95:359.

\bibitem{Nevot_RevPhysAppl_80}
Nevot, L., and P.~Croce. 1980.
\newblock Characterization of surfaces by grazing x-ray reflection -
  application to study of polishing of some silicate-glasses.
\newblock \emph{Rev. Phys. Appl.} 15:761--779.

\bibitem{Mattai_Biochemistry_89}
Mattai, J., H.~Hauser, R.~A. Demel, and G.~G. Shipley. 1989.
\newblock Interactions of metal ions with phosphatidylserine bilayer membranes:
  effect of hydrocarbon chain unsaturation.
\newblock \emph{Biochemistry}. 28:2322.

\bibitem{Daniel_Biointerphases_07}
Daniel, C., K.~E. Sohn, T.~E. Mates, E.~J. Kramer, J.~O. R\"adler, E.~Sackmann,
  B.~Nickel, and L.~Andruzzi. 2007.
\newblock Structural characterization of an elevated lipid bilayer obtained by
  stepwise functionalization of a self-assembled alkenyl silane film.
\newblock \emph{Biointerphases}. 2:109--118.

\bibitem{Fritz_JMB_09}
Fritz, G., C.~Koller, K.~Burdack, L.~Tetsch, I.~Haneburger, K.~Jung, and
  U.~Gerland. 2009.
\newblock Induction kinetics of a conditional {pH} stress response system in
  {{\em Escherichia coli}}.
\newblock \emph{J. Mol. Biol.} 393:272--286.

\bibitem{fischer2004}
Fischer, H., I.~Polikarpov, and A.~F. Craievich. {2004}.
\newblock {Average protein density is a molecular-weight-dependent function}.
\newblock \emph{Prot. Sci.} {13}:{2825--2828}.

\bibitem{Klauda_BiophysJ_06}
Klauda, J.~B., N.~Kucerka, B.~R. Brooks, R.~W. Pastor, and J.~F. Nagle. 2006.
\newblock Simulation-based methods for interpreting x-ray data from lipid
  bilayers.
\newblock \emph{Biophys. J.} 90:2796--2807.

\bibitem{Kucerka_CurrOpinColloidInterfaceSci_07}
Kucerka, N., M.~P. Nieh, J.~Pencer, T.~Harroun, and J.~Katsaras. 2007.
\newblock {The study of liposomes, lamellae and membranes using neutrons and
  X-rays}.
\newblock \emph{Curr. Opin. Colloid Interface Sci.} 12:17--22.

\bibitem{Kucerka_BiophysJ_08}
Kucerka, N., J.~F. Nagle, J.~N. Sachs, S.~E. Feller, J.~Pencer, A.~Jackson, and
  J.~Katsaras. 2008.
\newblock Lipid bilayer structure determined by the simultaneous analysis of
  neutron and x-ray scattering data.
\newblock \emph{Biophys. J.} 95:2356--2367.

\bibitem{Feigenson_Biochemistry_86}
Feigenson, G.~W. 1986.
\newblock On the nature of calcium-ion binding between phosphatidylserine
  lamellae.
\newblock \emph{Biochemistry}. 25:5819--5825.

\bibitem{Dietrich_BiophysJ_97}
Dietrich, C., R.~Merkel, and R.~Tampe. 1997.
\newblock Diffusion measurement of fluorescence-labeled amphiphilic molecules
  with a standard fluorescence microscope.
\newblock \emph{Biophys. J.} 172:1701--1710.

\bibitem{Hochrein_Langmuir_06}
Hochrein, M.~B., C.~Reich, B.~Krause, J.~O. R\"adler, and B.~Nickel. 2006.
\newblock Structure and mobility of lipid membranes on a thermoplastic
  substrate.
\newblock \emph{Langmuir}. 22:538--545.

\bibitem{Ladha_BiophysJ_96}
Ladha, S., A.~R. Mackie, L.~J. Harvey, D.~C. Clark, E.~J.~A. Lea,
  M.~Brullemans, and H.~Duclohier. 1996.
\newblock {Lateral diffusion in planar lipid bilayers: A fluorescence recovery
  after photobleaching investigation of its modulation by lipid composition,
  cholesterol, or alamethicin content and divalent cations}.
\newblock \emph{Biophys. J.} 71:1364--1373.

\bibitem{Liangfang_JChemPhys_05}
Liangfang, Z., and G.~Steve. 2005.
\newblock Lipid diffusion compared in outer and inner leaflets of planar
  supported bilayers.
\newblock \emph{J. Chem. Phys.} 123:211104.

\bibitem{Scomparin_EurPhysJ_09}
Scomparin, C., S.~Lecuyer, M.~Ferreira, T.~Charitat, and B.~Tinland. 2009.
\newblock {Diffusion in supported lipid bilayers: Influence of substrate and
  preparation technique on the internal dynamics}.
\newblock \emph{Eur. Phys. J. E}. 28:211--220.

\bibitem{WaismanM_MolCellBiochem_95}
Waisman, D.~M. 1995.
\newblock Annexin {II} tetramer: structure and function.
\newblock \emph{Mol. Cell Biochem.} 149-150:301--22.

\bibitem{Vacklin_Langmuir_05}
Vacklin, H.~P., F.~Tiberg, G.~Fragneto, and R.~K. Thomas. 2005.
\newblock Composition of supported model membranes determined by neutron
  reflection.
\newblock \emph{Langmuir}. 21:2827--2837.

\bibitem{Trauble_PNAS_74}
Trauble, H., and H.~Eibl. 1974.
\newblock Electrostatic effects on lipid phase-transitions - membrane structure
  and ionic environment.
\newblock \emph{Proc. Natl. Acad. Sci. USA}. 71:214--219.

\bibitem{Casal_Biochemistry_87}
Casal, H.~L., A.~Martin, H.~H. Mantsch, F.~Paltauf, and H.~Hauser. 1987.
\newblock {Infrared studies of fully hydrated unsaturated phosphatidylserine
  bilayers. Effect of lithium and calcium}.
\newblock \emph{Biochemistry}. 26:7395--7401.

\bibitem{Shreve_Langmuir_08}
Shreve, A.~P., M.~C. Howland, A.~R. Sapuri-Butti, T.~W. Allen, and A.~N.
  Parikh. 2008.
\newblock Evidence for leaflet-dependent redistribution of charged molecules in
  fluid supported phospholipid bilayers.
\newblock \emph{Langmuir}. 24:13250--13253.

\bibitem{Richter_Langmuir_04}
Richter, R., N.~Maury, and A.~Brisson. 2004.
\newblock On the effect of the solid support on the interleaflet distribution
  of lipids in supported lipid bilayers.
\newblock \emph{Langmuir}. 21:299--304.

\bibitem{Schalke_BiochimBiophysActa_00}
Schalke, M., P.~Kr\"uger, M.~Weygand, and M.~L\"osche. 2000.
\newblock {Submolecular organization of DMPA in surface monolayers: beyond the
  two-layer model}.
\newblock \emph{Biochim. Biophys. Acta}. 1464:113--26.

\bibitem{Watkins_PhysRevLett_09}
Watkins, E.~B., C.~E. Miller, D.~J. Mulder, T.~L. Kuhl, and J.~Majewski. 2009.
\newblock Structure and orientational texture of self-organizing lipid
  bilayers.
\newblock \emph{Phys. Rev. Lett.} 102:238101.

\bibitem{Lin_BiophysJ_06}
Lin, W.-C., C.~D. Blanchette, T.~V. Ratto, and M.~L. Longo. 2006.
\newblock Lipid asymmetry in {DLPC/DSPC}-supported lipid bilayers: a combined
  {AFM} and fluorescence microscopy study.
\newblock \emph{Biophys. J.} 90:228--37.

\bibitem{Ratto_BiophysJ_02}
Ratto, T.~V., and M.~L. Longo. 2002.
\newblock Obstructed diffusion in phase-separated supported lipid bilayers, a
  combined {AFM} and {FRAP} approach.
\newblock \emph{Biophys. J.} 83:3380--3392.

\bibitem{Fenz_Langmuir_09}
Fenz, S.~F., R.~Merkel, and K.~Sengupta. 2009.
\newblock {Diffusion and intermembrane distance: case study of avidin and
  E-cadherin mediated adhesion}.
\newblock \emph{Langmuir}. 25:1074--85.

\bibitem{Cremer_JPhysChemB_99}
Cremer, P.~S., and S.~G. Boxer. 1999.
\newblock Formation and preading of lipid bilayers on planar glass support.
\newblock \emph{J. Phys. Chem. B}. 103:2554--2559.

\bibitem{Richter_Langmuir_06}
Richter, R.~P., R.~Berat, and A.~R. Brisson. 2006.
\newblock {Formation of solid-supported lipid bilayers: An integrated view}.
\newblock \emph{Langmuir}. 22:3497--3505.

\bibitem{Dertinger_Langmuir_06}
Dertinger, T., I.~von~der Hocht, A.~Benda, M.~Hof, and J.~Enderlein. 2006.
\newblock Surface sticking and lateral diffusion of lipids in supported
  bilayers.
\newblock \emph{Langmuir}. 22:9339--9344.

\bibitem{gilmanshin1994}
Gilmanshin, R., C.~E. Creutz, and L.~K. Tamm. {1994}.
\newblock {Annexin-IV reduces the rate of lateral lipid diffusion and changes
  the fluid-phase structure of the lipid bilayer when it binds to negatively
  charged membranes in the presence of calcium}.
\newblock \emph{Biochemistry}. {33}:{8225--8232}.

\bibitem{Heimburg_BiophysJ_99}
Heimburg, T., B.~Angerstein, and D.~Marsh. 1999.
\newblock Binding of peripheral proteins to mixed lipid membranes: The effect
  of local demixing upon binding.
\newblock \emph{Biophys. J.} 76:2575--2586.

\bibitem{May_BiophysJ_00}
May, S., D.~Harries, and A.~Ben-Shaul. 2000.
\newblock Lipid demixing and protein-protein interactions in the adsorption of
  charged proteins on mixed membranes.
\newblock \emph{Biophys. J.} 79:1747--1760.

\bibitem{Dermine_JBiolChem_01}
Dermine, J.~F., S.~Duclos, J.~Garin, F.~St-Louis, S.~Rea, R.~G. Parton, and
  M.~Desjardins. 2001.
\newblock Flotillin-1-enriched lipid raft domains accumulate on maturing
  phagosomes.
\newblock \emph{J. Biol. Chem.} 276:18507--12.

\bibitem{faure2002}
Faure, A.~V., C.~Migne, G.~Devilliers, and J.~Ayala-Sanmartin. {2002}.
\newblock {Annexin 2 ``secretion{''} accompanying exocytosis of chromaffin
  cells: Possible mechanisms of annexin release}.
\newblock \emph{Exp. Cell Res.} {276}:{79--89}.

\bibitem{Esteban-Martin2009}
Esteban-Mart{\'\i}n, S., H.~J. Risselada, J.~Salgado, and S.~J. Marrink. 2009.
\newblock Stability of asymmetric lipid bilayers assessed by molecular dynamics
  simulations.
\newblock \emph{J. Am. Chem. Soc.} 131:15194--202.

\bibitem{Wallace2005}
Wallace, E.~J., N.~M. Hooper, and P.~D. Olmsted. 2005.
\newblock The kinetics of phase separation in asymmetric membranes.
\newblock \emph{Biophys. J.} 88:4072--83.

\end{thebibliography}

%
\clearpage
\section*{Tables}

\begin{table}[h]
\begin{center}
{\small {\sf
\begin{tabular}{|p{6cm}@{$\;\;$}|@{$\;\;$}c@{$\;\;$}|@{$\;\;$}c@{$\;\;$}|}
\hline
Number of $\mathsf{e^-}$  & 25 \% POPS + 75 \% POPC & 50 \% POPS + 50 \% POPC \\
\hline
headgroup (phosphate + $\mathsf{Ca^{2+}}$ + serine/choline)  & 98.75 & 100.5 \\
backbone (glycerol + carbonyl group) & 67 & 67 \\
chain (two alkyl groups) & 238 & 238\\
chain ends (four methyl groups [two per opposite lipid]) & 36  & 36\\
\hline
\end{tabular}
}
}
\caption{\label{TABinputParameters} Input parameters from the chemical structures as used for the decomposition of the electron density profiles of the bilayer (see text).
}
\end{center}
\vspace*{0.2cm}
\end{table}

\begin{table}[h]

{\small {\sf
$\,$\\
(a)

$\,$\\
\begin{tabular}{|p{5cm}@{$\;\;$}|@{$\;\;$}c@{$\;\;$}|@{$\;\;$}c@{$\;\;$}|}
\hline
& 25 \% POPS + 75 \% POPC & 50 \% POPS + 50 \% POPC \\
\hline
head-to-head distance $\mathsf{[\AA]}$ & $\mathsf{44\pm1}$ & $\mathsf{42\pm1}$ \\
area per lipid $\mathsf{[\AA^2/lipid]}$ & proximal $\mathsf{69\pm3}$; distal $\mathsf{60\pm2}$ & proximal $\mathsf{67\pm2}$; distal $\mathsf{51\pm3}$ \\
\hline
\end{tabular}
$\,$\\

(b)

$\,$\\
\begin{tabular}{|p{5cm}@{$\;\;$}|@{$\;\;$}c@{$\;\;$}|@{$\;\;$}c@{$\;\;$}|}
\hline
& 25 \% POPS + 75 \% POPC & 50 \% POPS + 50 \% POPC \\
\hline
Anx A2t thickness $\mathsf{[\AA]}$ & $\mathsf{59\pm6}$ & $\mathsf{67\pm5}$ \\
head-to-head distance $\mathsf{[\AA]}$ & $\mathsf{48\pm3}$ & $\mathsf{45\pm3}$ \\
area per lipid $\mathsf{[\AA^2/lipid]}$ & proximal $\mathsf{65\pm5}$; distal $\mathsf{46\pm3}$ & proximal $\mathsf{59\pm4}$; distal $\mathsf{41\pm4}$ \\
\hline
\end{tabular}
}
}
\caption{\label{TABbilayerParameters} Bilayer parameters of SLBs in calcium buffer extracted from XR measurements  ({\it a}) before and ({\it b}) after Anx A2t incubation. Proximal and distal indicate the SLB leaflets facing the silicon surface and the buffer, respectively. Errors were estimated as described in {\it Materials \& Methods}.
}

\vspace*{0.2cm}
\end{table}

\clearpage
\section*{Figure Legends}
\subsubsection*{Figure~\ref{FIGscheme}.}
 Cartoon of the proposed conformations of Anx A2t when bound to a single bilayer ({\it a}) and the resulting conformations when binding to two bilayers ({\it b}). The Anx A2 monomers are represented by half spheres while the p11 (S100A10) dimers are represented by ellipsoids. The side-by-side configuration ({\it a, top}) is formed by binding of both Anx A2 monomers to one bilayer with the S100A10 dimer on top. From this conformation it was proposed that an octameric complex might be formed in the presence of a second bilayer ({\it b, top}).  The vertical conformation ({\it b, bottom}) is formed by one Anx A2 monomer binding to one bilayer and the second A2 monomer binding to a second membrane. From this model the same, albeit hypothetical, model was proposed for Anx A2t binding to a single bilayer Fig.~\ref{FIGscheme}~({\it a, bottom}).

\subsubsection*{Figure~\ref{FIGreflectivity}.}
X-ray reflectivity data of a 25 mol \% POPS containing phospholipid membrane ({\it upper data points}) and a 50 mol \% POPS containing phospholipid bilayer ({\it lower data points})  in calcium buffer ({\it squares}) and after incubation with Anx A2t ({\it triangles}). The ellipsoids  highlight the change in reflectivity signal due to Anx A2t binding. The solid lines represents the 15 best fit obtained by Parrats algorithm, see {\it Materials and Methods} for all details. For better comparison data was multiplied by $q^{4}$ and shifted on the y-axis.

\subsubsection*{Figure~\ref{FIGelectronDensity}.}
Electron density profiles of a 25 mol \% POPS ({\it a}) and a 50 mol \% POPS ({\it b}) containing SLB in calcium buffer, as obtained by quantitative fits of the data in Fig.~\ref{FIGreflectivity}. The inset shows the individual profiles obtained by the 15 best fits ({\it grey lines}) as well as their arithmetic mean ({\it black lines}). Blue, grey, and light grey areas indicate contributions from buffer, bilayer, and silicon support, respectively. These contributions add up to the total profile ({\it red line}), which is superimposed on the arithmetic mean ({\it black line}). The contribution from the bilayer can be further separated into contributions from both leaflets ({\it dark grey areas}). Each leaflet is separated into its head, its alkyl chain part, and its chain termini ({\it green}, {\it dark green}, and {\it yellow area}, respectively). Individual contributions of the head are shown by turquoise lines (phosphate plus choline/serine group) and green lines (glycerol plus carbonyl group).

\subsubsection*{Figure~\ref{FIGelectronDensity_w_prot}.}
Electron density profiles of POPS containing SLBs after Anx A2t incubation.
The electron density profile of a 25 mol \% POPS ({\it a}) and a 50 mol \% POPS containing SLB ({\it b}) is shown  after incubation with Anx A2t. The color code is identical to Fig.~\ref{FIGelectronDensity}. Additionally, the contribution of Anx A2t to the electron density profile is represented by the red area. Individual contributions to the Anx A2t signal can be assigned to the Anx A2 monomer ({\it dark red areas}) and S100A10 dimer ({\it thin red line}), for details see text.

\subsubsection*{Figure~\ref{FIGbleaching}.}
Continuous bleaching series of a 25 mol \% POPS containing bilayer before ({\it a}) and after ({\it b}) Anx A2t incubation. Images were taken before bleaching, after 60\,s and 120\,s of bleaching time. The light intensities were chosen similar in ({\it a}) and ({\it b}), such that the bleaching rates were comparable in both cases. The white rim at the border of the aperture indicates fresh non-bleached fluorescent lipids diffusing into the bleached area. The white lines show the radially averaged fluorescence profiles of the images. The arrows mark the decay-width of the fluorescence signal at the rim of the aperture: the reduced width after Anx A2t incubation in ({\it b}) is indicative of a reduced lipid mobility.

\subsubsection*{Figure~\ref{FIGdiffusionConstants}.}
Diffusion constant of a pure POPC, a 25 mol \% POPS and a 50 mol \% POPS bilayer in calcium buffer ({\it dark grey bars}) and after Anx A2t incubation ({\it light grey bars}). Values are mean values from at least 3 independent samples with at least 5 independent measurement points on each sample. Error bars represent standard deviations from the mean.

\begin{figure}
\centerline{\includegraphics[width=6in]{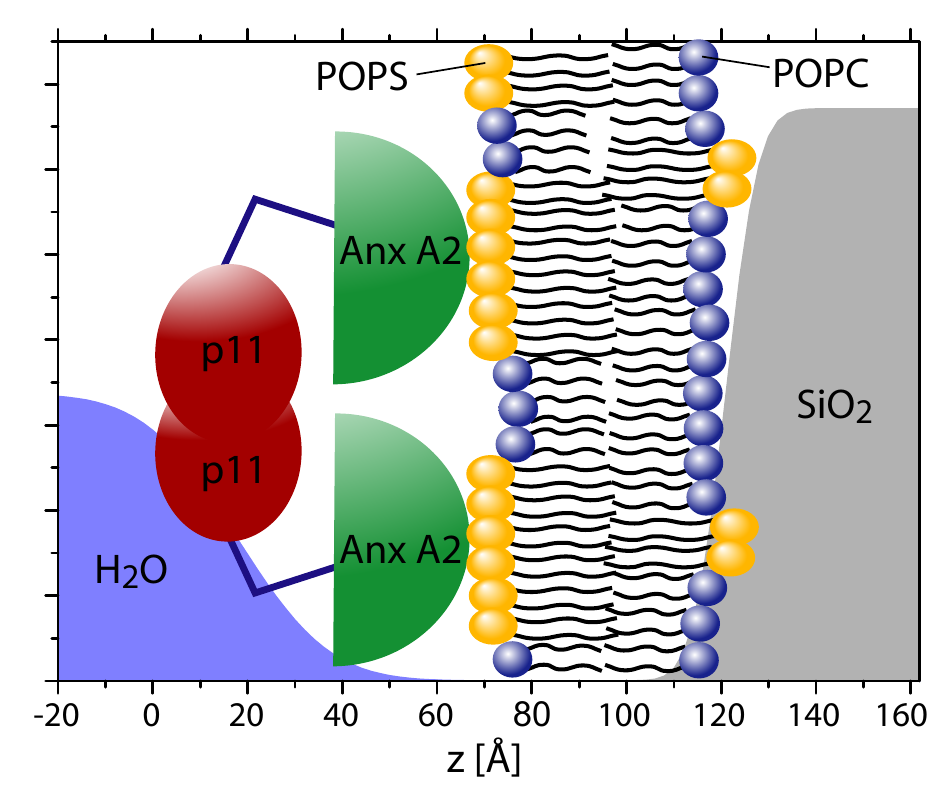}}
\caption{\label{FIGsummary} Fritz et al. }
\end{figure}

\subsubsection*{Figure~\ref{FIGsummary}.}
Proposed quarternary structure of Anx A2t upon binding to a supported lipid bilayer. Two Anx A2 monomers ({\it half spheres}) bind to the distal leaflet of the membrane while a p11 dimer ({\it ellipsoids}) sits on top of this structure. The presence of calcium and Anx A2t lead to a densification of the protein-facing leaflet of the bilayer, presumably due to an enrichment of POPS as compared to POPC.

\clearpage

\end{document}